\documentstyle[aps,twocolumn]{revtex}
\input epsf
\tightenlines
\begin{document}
\draft
\preprint{VECC/NTH97xxx}
\title{\bf  Radiative Energy-Loss of Heavy Quarks in a Quark-Gluon Plasma}
\author{Munshi Golam Mustafa}
\address{Saha Institute of Nuclear Physics, 1/AF Bidhan Nagar, Calcutta
700 064, India}
\author{Dipali Pal and Dinesh Kumar Srivastava}
\address{Variable Energy Cyclotron Centre, 1/AF Bidhan Nagar, Calcutta 
700 064, India}
\author{Markus Thoma}
\address{Institut f{\"u}r Theoretische Physik, Universit{\"a}t Giessen, 
D-35392 Giessen, Germany}
\date{\today}
\maketitle

\begin{abstract}
We estimate the radiative energy-loss of heavy quarks, produced from 
the initial fusion of partons, while propagating in a quark-gluon 
plasma which may be formed in the wake of relativistic heavy ion 
collisions. We find that the radiative energy-loss for heavy quarks
is larger than the collisional energy-loss for all energies. We point out the
consequences on possible signals of the quark-gluon plasma.
\end{abstract}
%
%
\narrowtext

\vskip 0.5in

One of the most interesting predictions of QCD
is the transition from the confined/chirally broken phase to the 
deconfined/chirally
symmetric state of quasi-free quarks and gluons, the so-called quark-gluon
 plasma (QGP).
Relativistic heavy ion collisions are being studied with the intention of 
investigating the properties of the QGP~\cite{hm96}.
While experiments at AGS and SPS 
continue, new experiments have been planned at RHIC and LHC  
with centre of mass energies 200 AGeV and 5.5 ATeV, respectively.
 During the past
decade many different signatures of the transition to the QGP have been proposed. 
A promising example is the emission of 
penetrating probes such as dileptons and single photons, which can reveal 
the early parton dynamics and the history of evolution of the plasma.
 Similarly, the
production and propagation of open charm and high energy jets 
in a dense medium, can provide information 
\cite{xnw} about parton scattering and thermalization of the partonic system. 
Jets are expected to show up at collider energies at RHIC and LHC. 
 
Heavy quark pairs are mostly produced from the initial fusion of
partons (mostly from $gg\rightarrow Q{\bar Q}$, but also from $q{\bar q}
\rightarrow Q{\bar Q}$, where $q$ denotes one of the lighter quarks and 
$Q$ is a heavy quark) of the colliding nucleons and also from the QGP, if the 
initial temperature is high enough, which is likely to be achieved 
at RHIC and LHC energies. The charm quarks will be produced on a time 
scale of $1/2m_c\simeq$ 0.07 fm/c, which would be as low as
$\simeq$ 0.02 fm/c for bottom quarks.
 There is no production of heavy quarks at late times in the QGP 
and none in the hadronic matter. Thus, the total number of heavy quarks
gets frozen very early in the history of the collision which makes them a 
good candidate for a probe of the QGP.  Immediately upon their production, 
these heavy quarks will propagate through the deconfined matter and start 
losing energy. The energy-loss suffered by these quarks will determine the
shape of the dilepton spectra produced from correlated charm (or bottom) decay
which provides a large background to dilepton production from annihilation of
quarks in the plasma. We shall come back to this aspect towards the
end of this Letter.

There are two contributions to the energy-loss of a heavy quark in 
the QGP: one caused by elastic collisions with the light partons of the 
QGP and the other by radiation of the decelerated color charge, 
$i.e.$, bremsstrahlung of gluons. 
There is an extensive body of 
literature~ \cite{bjor,ben,thoma,mrow,koike,thoma1} 
on the collisional energy-loss of energetic quarks considering elastic
 collisions
with the quarks and gluons ($Qg\rightarrow Qg$ and $Qq\rightarrow Qq$)
of the dense medium. A complete leading order result for the collisional 
energy-loss of heavy quarks has been found using the hard thermal loop
resummation technique \cite{braaten}. 

It is well known that the contribution of the radiative processes
($Qq\rightarrow Qqg$ and $Qg\rightarrow Qgg$) is of the same order
in the coupling constant as the
collisional energy-loss \cite{braaten}.  The estimate of the radiative 
energy-loss in the past has been discussed by a number of 
authors~\cite{gyul3,gyul1,gyul2,baier} within perturbative QCD 
taking into account the Landau-Pomeranchuk suppression due to 
multiple collisions. These studies, however, were limited to the case of  
massless energetic quarks and gluons.
As far as we know, there is no estimate of the radiative energy-loss 
for heavy quarks in the literature.
It is not easy to extend the sophisticated treatment of multiple
scattering formulated by the authors of Ref.~\cite{baier}
to the case of heavy quarks. 

Until such a detailed investigation is performed, an extension of the
work of Ref.~\cite{gyul3} for the radiative energy-loss of massless quarks
to the case of heavy quarks can provide valuable insight. 
This approach is similar to the one by Gyulassy, Wang
and Pl\"umer~\cite{gyul2} and leads to almost identical
results in the case of light partons. Baier et al.~\cite{baier},
on the other hand, found a different dependence of the
radiative energy-loss on the energy of the parton
by including the rescattering of the emitted gluon in
the QGP. However, inserting typical values for the energy
of the parton, the temperature, and the coupling constant
yields quantitatively similar results. Therefore
we will restrict ourselves to the simple approach of 
Ref.~\cite{gyul3} in the present work and discuss
the consequences of our estimate for signatures of the QGP.

\begin{figure}
\epsfxsize=3.75in
\vskip -0.6in
\epsfbox{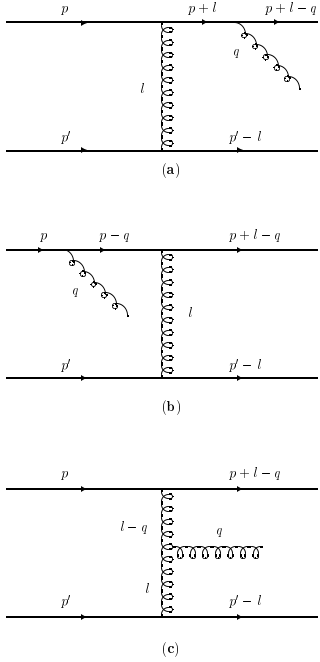}
\vskip -1in
\caption{Feynmann diagrams for gluon bremsstrahlung from quarks.}
\end{figure}

We start from an expression for the gluon emission probability
which has been derived by Gunion and Bertsch~\cite{gunion} in the
case of light partons assuming a factorization of the matrix elements 
of Fig.~1 into elastic scattering and gluon emission. It can be shown
that their result also holds if one of the light quarks in Fig.~1
is replaced by a heavy quark assuming that the engery of the emitted
gluon is not too large ($q_0\ll \sqrt{s}$). The final result for the
multiplicity distribution of the radiated gluon can then be written as
\begin{equation}
\frac{dn_g}{d\eta d^2q_{\bot}} = \frac{C_A\alpha_s}{\pi^2} 
\frac{l_{\bot}^2}{q_{\bot}^2 \left( {\vec q}_{\bot} - {\vec l}_{\bot}
\right )^2} \ \  , \label{mdist}
\end{equation}
where $q = (q_0,{\vec q}_{\bot},q_3)$ and $l=(l_0,{\vec l_{\bot}},l_3)$ are the four 
momenta of the emitted and the exchanged gluons, respectively, and $\eta= 
(1/2)\ln[(q_0+q_3)/(q_0-q_3)]$ is the rapidity. $C_A=3$ is the 
Casimir invariant of the adjoint representation. 
The factorization in (\ref{mdist}) was obtained in the limit
$x\,l_{\bot} << q_{\bot}$, where $x$ is the fractional  
momentum carried by the radiated gluon relative 
to the maximum available, such that the radiation is confined 
to a uniform (central) rapidity region, the most important
zone in relativistic heavy ion collisions. 
Eq.(\ref{mdist}) holds also for gluon bremsstrahlung emitted by a gluon.

Now we can estimate the radiative energy-loss per 
unit length for heavy quarks by multiplying the interaction rate 
$\Gamma$ and the average energy-loss per
collision $\nu$, which is given by the average of the probability 
of radiating a
gluon times the energy of the gluon. One can further correct for the 
Landau-Pomeranchuk effect by including a 
formation time restriction \cite{gyul3,gyul1} through a step function
$\theta(\tau-\tau_f)$. This puts a restriction
on the phase space of the emitted gluons in which the formation time, 
$\tau_f$ must be smaller than the interaction time, $\tau=1/\Gamma$.
The formation time is estimated by requiring the separation between
the emitted gluon and the parton from which it is emitted to be
$r_{\bot}=v_{\bot} t>1/q_{\bot}$ ($q_{\bot}\equiv |{\vec q}_{\bot}|$)
according to the uncertainty principle.
Using $v_{\bot}=q_{\bot}/q_0$ and $q_0=q_{\bot}\mathrm{cosh} \eta$,
we find $\tau _f={\mathrm{cosh}} \eta /q_{\bot}$.

The average radiative energy-loss per collision is calculated as
\begin {equation}
\nu=\langle \ n_g q_0 \ \rangle=\int d\eta d^2q_{\bot} \frac{dn_g}
{d\eta d^2q_{\bot}} \ q_0 \ \theta\left (\tau-\tau_f\right ) \ .
\label{ave}
\end{equation}
Performing the integration in (\ref{ave}) in the limit 
$(q_{\bot}\tau)^2 > > 1$ and $q_{\bot} >> l_{\bot}$, we get
\begin{equation}
\nu\simeq \frac{6\alpha_s}{\pi} \langle \ l_{\bot}^2 \ \rangle \tau 
\ln \left (\frac{q_{\bot}^{\mathrm {max}}}{q_{\bot}^{\mathrm{min}}}
\right ) \ . 
\label{ave1}
\end{equation} 
For the infrared cut-off $q_{\bot}^{\mathrm{min}}$ we choose the Debye 
screening mass of a pure gluon gas,
\begin{equation}
q_{\bot}^{\mathrm{min}}=\mu_D = \sqrt{4\pi\alpha_s}\, T \ , \label{dmass}
\end{equation}
where $T$ is the temperature of the system.
For heavy quarks of mass $M$, the square of the 
maximum transverse momentum of the emitted gluon is given by
\begin{equation}
 (q_{\bot}^{\mathrm {max}})^2 = {\left \langle  \frac {(s - M^2)^2}{4s}  
\right \rangle } \ \label{qtmax},
\end{equation}
where $s$ is the Mandelstam variable.  To evaluate (\ref{qtmax})
we need to compute $\langle s\rangle $ and $\langle 1/s \rangle$ 
leading to 
\begin{eqnarray}
& \langle s  \rangle & \ = \ M^2 \ + \ 2 p' E \ , \nonumber \\
& \langle 1/s \rangle & \ = \  
 \ \frac {1}{4p'p} \ln \left [ \frac {M^2+2E p' +2p p'
}{M^2+2E p' -2p p' }\right ] \ , \label{avgs}
\end{eqnarray}
where $E$ and $p$ are the energy and momentum of a incoming heavy 
quark and $p'$ is the average momentum of the light quark or gluon of the 
QGP. The average value of $p'$ can be taken as $\sim 3T$. 
Now, (\ref{qtmax}) becomes,
\begin{equation}
 (q_{\bot}^{\mathrm {max}})^2 = \frac{3ET}{2} - \frac {M^2}{4}
  + \frac {M^4}{48pT} \ln \left [ \frac {M^2+6ET+6pT}{M^2+6ET-6pT} \right ]
 .  \label{qtmax1}
\end{equation}
The average momentum transfer of the scattering process is defined as
\begin{equation}
\langle l_{\bot}^2 \rangle \simeq \langle l^2 \rangle \equiv 
\langle t \rangle
= \frac {\int_{\mu_D^2}^{q_{\bot}^{\mathrm{max}^2}} dt \ t \ d\sigma/dt} 
 {\int_{\mu_D^2}^{q_{\bot}^{\mathrm{max}^2}}\ dt \ d\sigma/dt} \ , 
\label{avet}
\end{equation}
where the differential cross section for elastic scattering is
\begin{equation}
\frac{d\sigma}{dt} = \frac{1}{16\pi\left ( s -M^2 \right ) ^2} 
|{\cal M}|^2 \ , \label{difc}
\end{equation}
and the square of the matrix element, $|{\cal M}|^2$ can 
be obtained from Ref.~\cite{comb}.
In the limit $|t|<< s$, the differential cross section in 
(\ref{difc}) can be approximated by
\begin{equation}
\frac{d\sigma}{dt} \sim \frac{1}{t^2} \ \ . \label{adifc}
\end{equation}
We have checked that the modification of the energy-loss using the full 
expression for $|{\cal M}|^2$ is negligible (see below).
Combining (\ref{avet}) to (\ref{adifc}) we get
\begin{equation}
\langle l_{\bot}^2 \rangle \simeq  \frac{\mu_D^2 \left 
(q_{\bot}^{\mathrm{max}}
\right )^2} { \left (q_{\bot}^{\mathrm{max}}\right )^2 -\mu_D^2 } \ \ln 
\left [ \frac {\left (q_{\bot}^{\mathrm{max}}\right )^2}{\mu_D^2} \right ] \ .
\label{avlt}
\end{equation}
The radiative energy-loss for heavy quarks is then obtained by
combining (\ref{ave1}) and (\ref{avlt}) and multiplying by $\Gamma 
= 1/\tau$, 
\begin{equation}
\left (- \frac{dE}{dx}\right )_{\mathrm{rad}}= \frac{3\alpha_s}{\pi} 
\frac{\mu_D^2 \left (q_{\bot}^{\mathrm {max}} \right)^2 } 
{\left (q_{\bot}^{\mathrm {max}}\right )^2 \! - \! \mu_D^2} \ln^2 \left [
\frac{\left (q_{\bot}^{\mathrm {max}} \right )^2}{\mu_D^2}\right ] . 
\label{rad}  
\end{equation}

Since the mass of the quark in this expressions enters only via the maximum
transverse momentum (\ref{qtmax1}) the radiative energy-loss of a heavy quark
differs from the one of a massless quark only for small energies of the order
of $M$. 

Let us also recall the expression for the 
collisional energy-loss of heavy quarks considered 
in Ref.~\cite{braaten} using the hard thermal loop resummation technique. In the
domain $E<<M^2/T$, it reads
\begin{eqnarray}
\left (-\frac{dE}{dx}\right )_{\mathrm{coll.}} &=& \frac{8\pi\alpha_s^2 T^2}
{3} \left (1+\frac{n_f}{6}\right ) \left [ \frac{1}{v} - \frac{1-v^2}
{2v^2} \right.\nonumber \\
&\times& \left.\ln \left (\frac {1+v}{1-v}\right ) \right ] \ln \left [
2^{\frac{n_f}{6+n_f}}B(v)\frac{ET}{m_gM}\right ] , \nonumber \\
\label{coll1}
\end{eqnarray}
whereas for $E>>M^2/T$, it is 
\begin{eqnarray}
\left (-\frac{dE}{dx}\right )_{\mathrm{coll.}} &=& \frac{8\pi\alpha_s^2 T^2}
{3} \left ( 1+\frac{n_f}{6}\right ) \nonumber \\
&\times& \ln \left [ 2^{\frac{n_f}{2(6+n_f)}}0.92\frac{\sqrt{ET}}{m_g}\right ] , 
\label{coll2}
\end{eqnarray}
where $v$ is the velocity of the heavy quarks, $B(v)$ is a smooth function
of $v$, which can be taken approximately as $0.7$, $n_f$ is the number of light 
quark flavours taken as 2.5, and $m_g=\sqrt {(1+n_f/6)/3}g_sT$ the thermal gluon mass.

It should be noted that the collisional and the radiative energy-loss
 are of the same
order in the coupling constant~\cite{braaten}, although the latter 
is caused by higher order
diagrams within naive perturbation theory. 
The reason for this behaviour is the fact
that the interaction rate entering into the radiative energy-loss suffers from a
quadratically infrared singularity using a bare propagator for the 
exchanged gluon, 
whereas this divergence is reduced to a logarithmic one for the 
collisional energy-loss. 
This reduction is caused by the presence of the energy transfer of the exchanged
gluon in the definition of the collisional energy-loss~\cite{braaten}. 
In the case of the
radiative energy-loss, on the other hand, this factor is absent, 
because the energy-loss
is caused by the emitted and not by the exchanged gluon. 
Using a hard thermal loop
resummed propagator the quadratic singularity in the interaction rate is 
reduced to a logarithmic
one and the final result is of higher order ($\Gamma \sim \alpha _s$) 
than naively expected.
Multiplying this rate by the gluon emission probability (\ref{mdist}) yields the
result (\ref{rad}) of order $\alpha _s^2$. Using a resummed propagator for the
 collisional energy-loss 
leads to the finite expressions (\ref{coll1}) and (\ref{coll2}) of 
the same order.

\begin{figure}
\epsfxsize=3.25in
\epsfbox{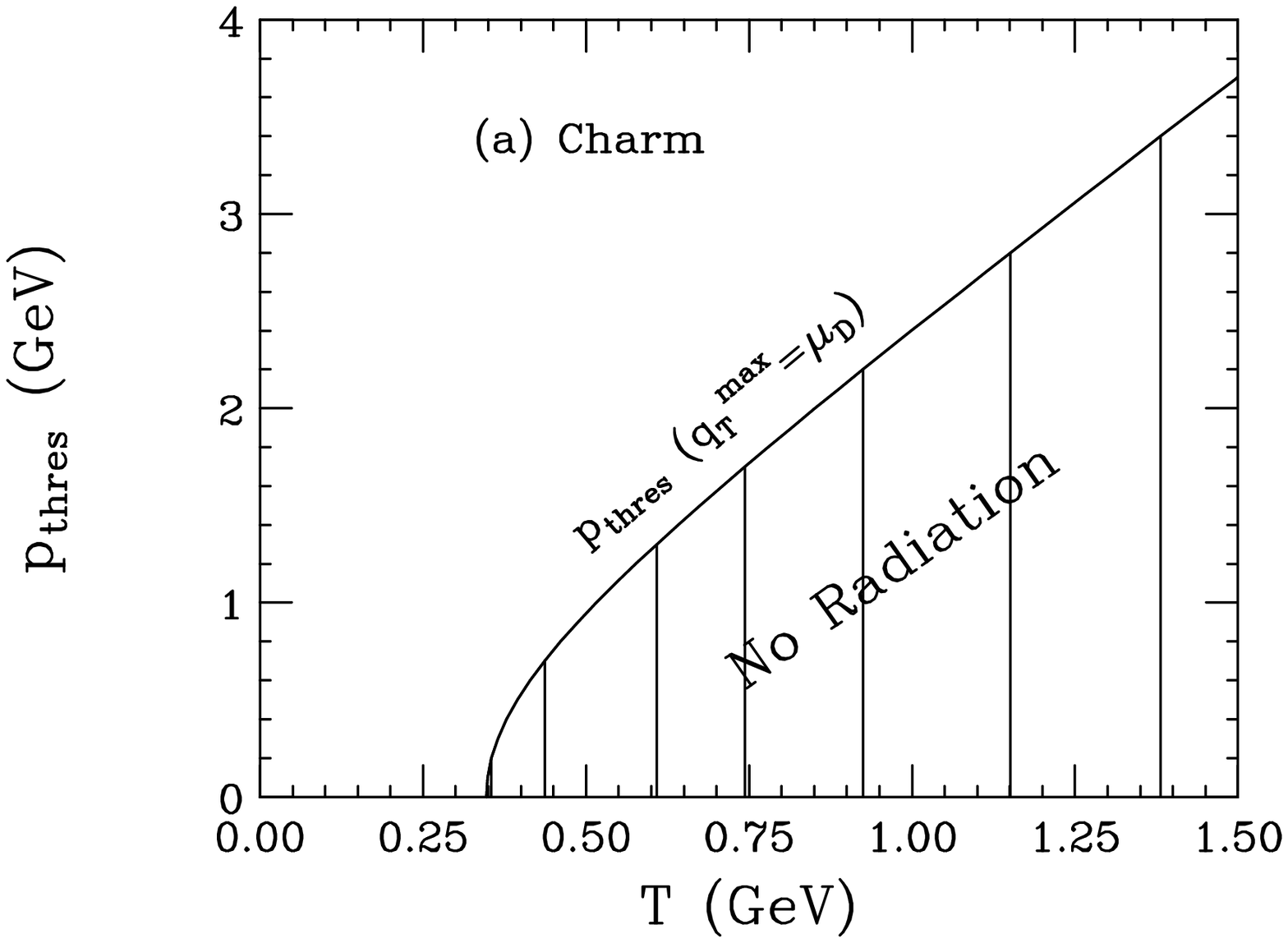}
\vskip 0.15 in
\epsfxsize=3.25in
\epsfbox{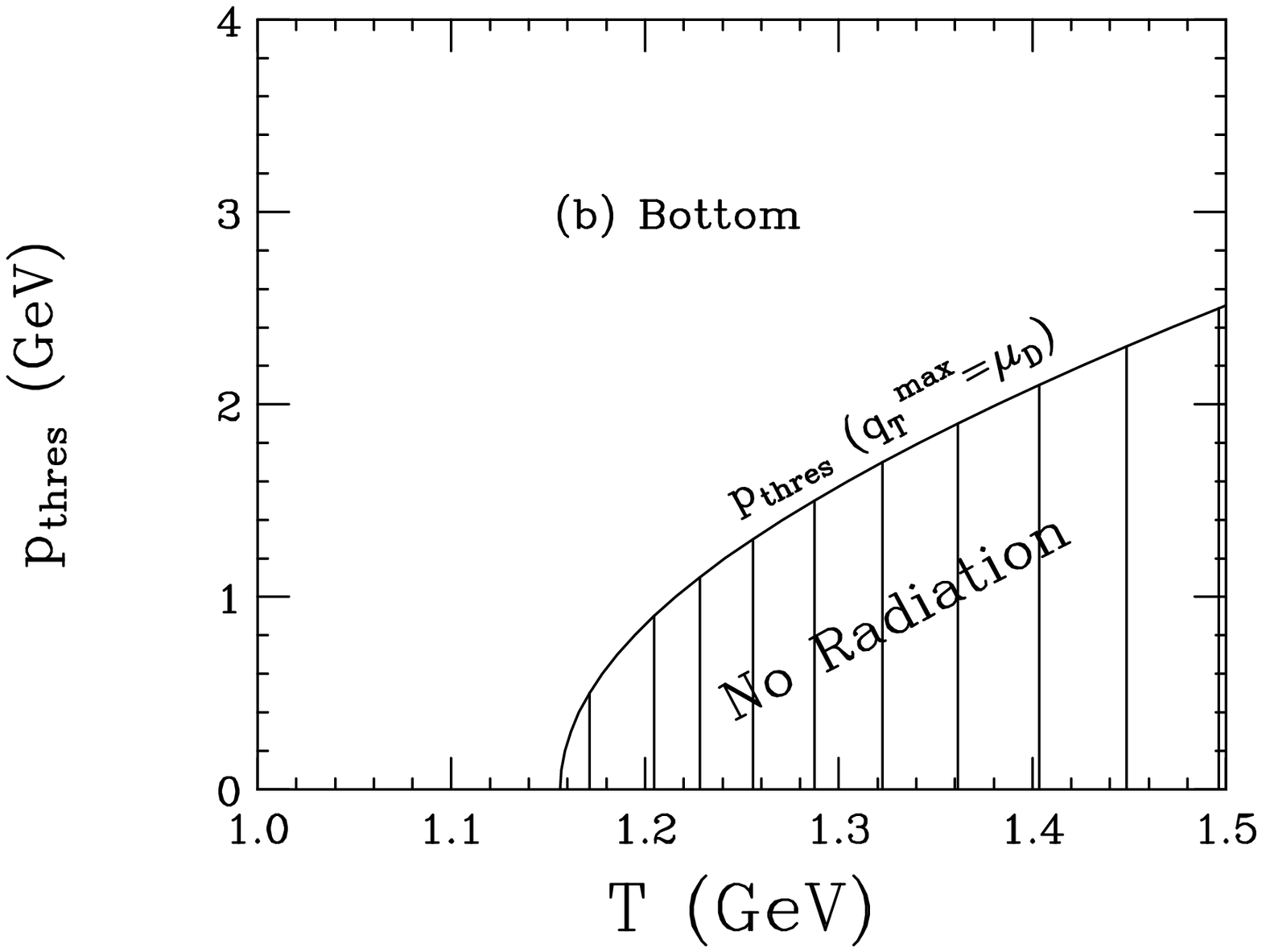}
\vskip 0.2 in
\caption{The threshold momenta of the heavy quark 
as a function of $T$.}
\end{figure}

The expression for the radiative energy-loss in (\ref{rad}) exhibits 
a threshold behaviour: for ${(q_{\bot}^{\mathrm {max}})^2} < \mu_D^2$ 
there is no radiative energy-loss which is shown by the hatched area in Fig.~2.
(Obviously, this behaviour will be different for mass-less quarks.)
We see that the value of the threshold momenta of the heavy quarks, 
below which there is no radiation, increases with increasing 
temperature.

In Fig.~3, we compare our results with that of the collisional energy-loss 
for heavy quarks obtained in Ref.~\cite{braaten} as 
a function of energy at a temperature $T$ = 500 MeV and $\alpha_s$ =0.3
for charm and bottom quarks. Before discussing our results, we give 
the justification for the approximation made in (\ref{adifc})
for the differential 
cross section used in computing the radiative energy-loss,
which enabled us to obtain the closed form given above. 
The solid lines 
represent the radiative energy-loss of heavy quarks with full 
$|{\cal M}|^2$ whereas the dashed lines correspond to that 
with the approximate expression. We see that retaining only the
$\sim 1/t^2$ term in $d\sigma/dt$ is sufficient for our purpose.
The dash-dotted lines in Fig.~3
represent the collisional energy-loss obtained in Ref.~\cite{braaten}.
We find that the radiative energy-loss dominates over the collisional one at
all energies. For $E>20$ GeV the difference amounts to an order of magnitude. 

\begin{figure}
\epsfxsize=3.25in
\epsfbox{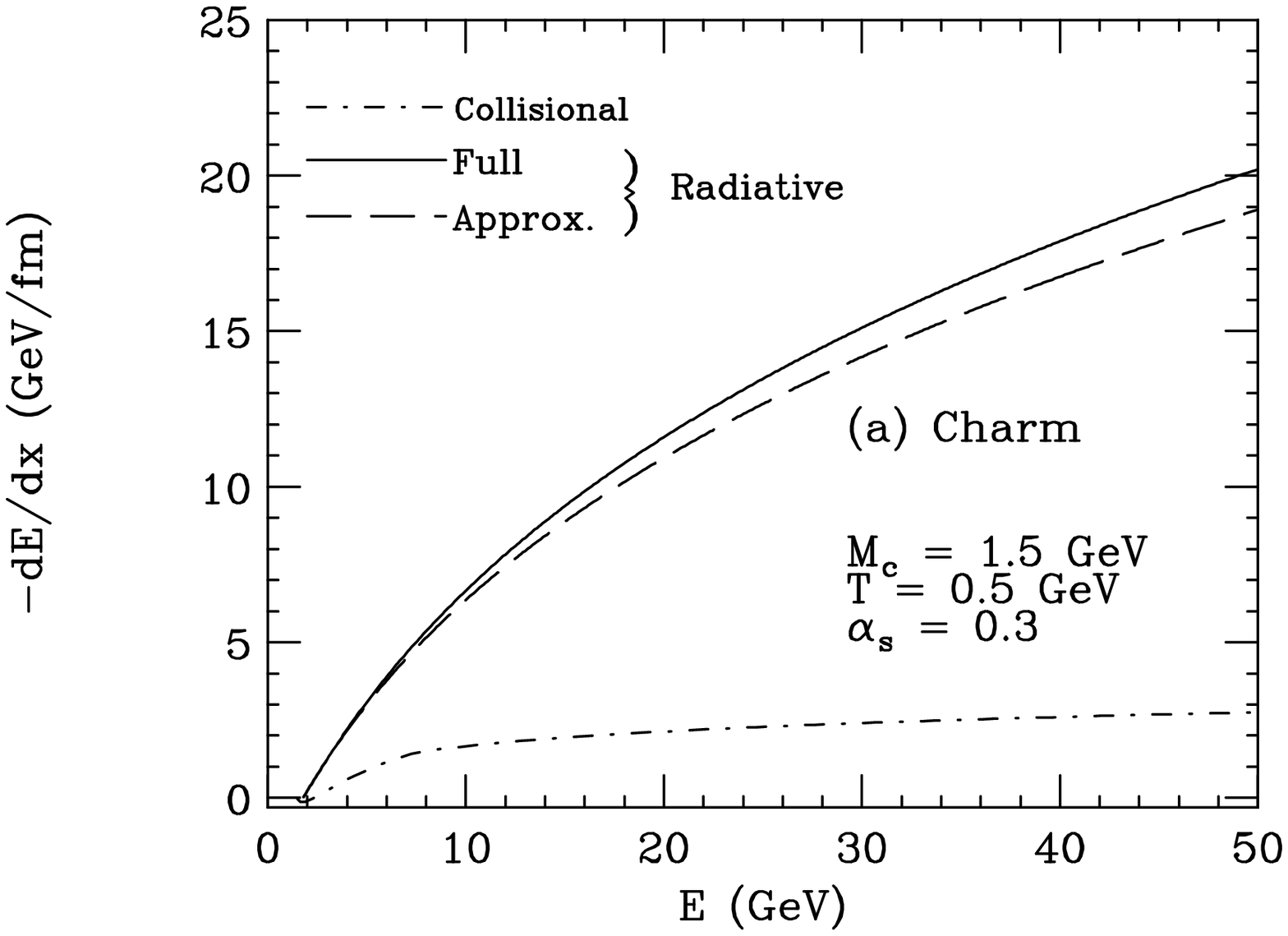}
\vskip 0.15 in
\epsfxsize=3.25in
\epsfbox{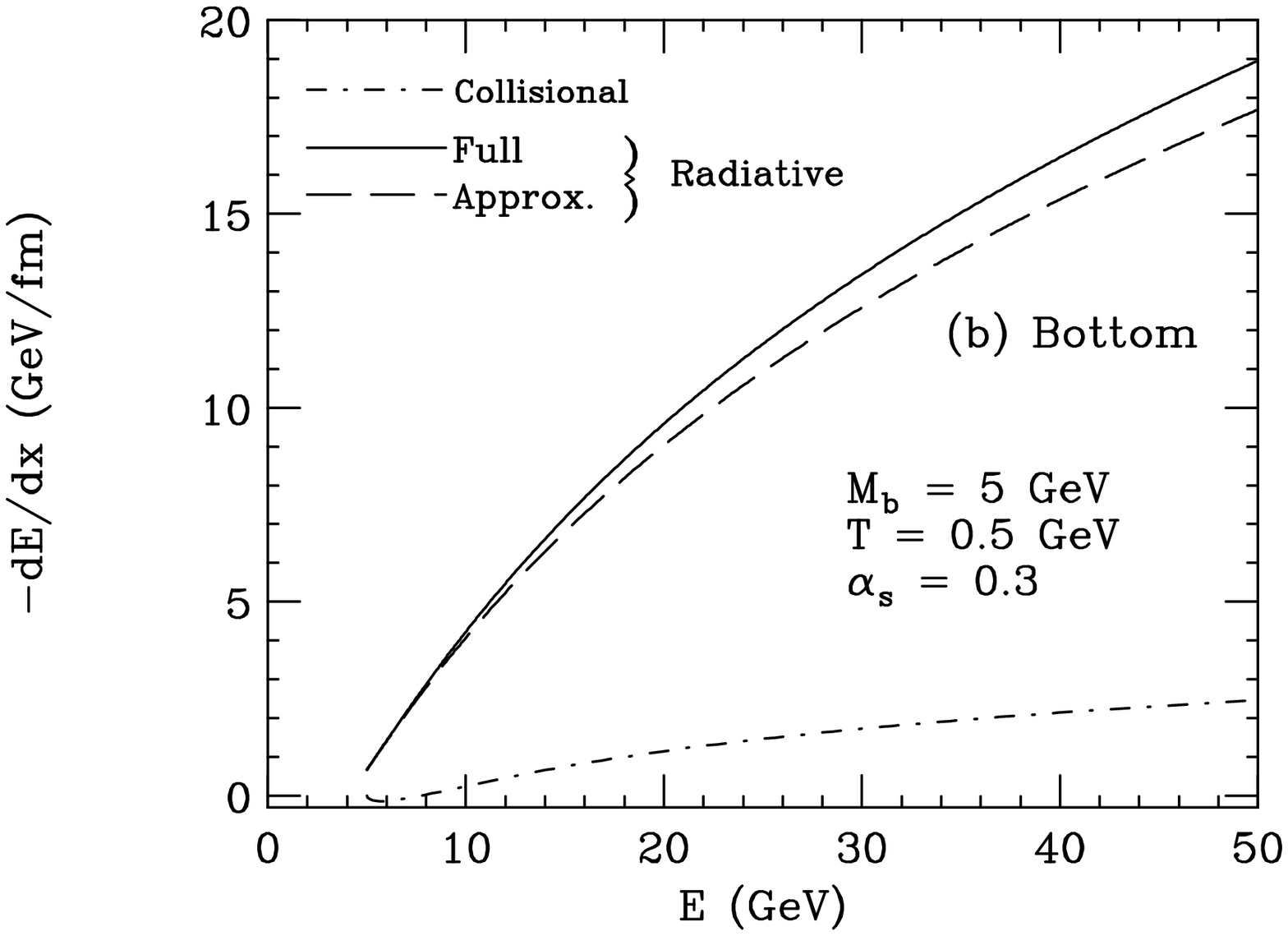}
\vskip 0.2 in
\caption{The energy-loss of heavy quarks as a function of their energy.}
\end{figure}

The QGP expected to be produced at RHIC and LHC is likely to be
far from chemical equilibrium, initially. Chemical reactions among the
partons will then push it towards a chemical 
equilibrium~\cite{klaus,biro,munshi2}. The evolution of the
temperature and the quark and gluon fugacities\footnote{The fugacities are 
defined~\cite{biro} as $f_i=\lambda_i \tilde{f}_i$,
where $f$ is the distribution of the partons and $\tilde{f}$ is the 
corresponding equilibrium distribution.} at RHIC and LHC eneriges
has recently been obtained~\cite{munshi2} for such a scenario with
initial conditions from a Self Screened Parton Cascade
Model~\cite{sspc}. As a first estimate, the expressions for the collisional
energy-loss given
earlier (Eqs.~(\ref{coll1}), (\ref{coll2})) can be modified 
by replacing the terms $(1+ n_f/6)$  by $(\lambda_g+\lambda_q n_f/6)$ and 
$n_f/(6+n_f)$ by $\lambda_q n_f/(6 \lambda_g+\lambda_q n_f)$ to account
for the departure from chemical equilibrium. Alternatively
one may use the results of Ref.~\cite{munshi} for a non-equilibrium
plasma. The radiative energy-loss is modified by using the non-equilibrium
Debye mass~\cite{biro}
\begin{equation}
\mu_D^2=4\pi\lambda_g \alpha_s T^2.
\end{equation}
in (\ref{rad}).

This has interesting consequences. It has been shown recently~\cite{munshi} 
that considering {\em only}
the collisional energy-loss, in this manner, amounts to having only a 
 small drag
on the motion of heavy quarks in such a plasma, at least at RHIC 
energies, where the charm quarks were found to loose only $\sim$ 10\% of
their energy  during their propagation, through the plasma and upto
40\% of their initial energy at LHC, in a collision involving two gold
nuclei~\cite{correction}.

The drag acting on the heavy quark is conveniently defined by
 writing
\begin{equation}
-\frac{dE}{dx}= A p,
\end{equation}
where $A$ denotes a drag-coefficient in the spirit of the treatment used
earlier in literature~\cite{ben,munshi} and $p$ is the momentum of the 
heavy quark. Adding the collisional and radiative energy-loss experienced by a heavy quark
in such an equilibrating  and cooling plasma,
we have found that
\begin{equation}
A\simeq C/\tau,
\end{equation}
where $C$ is a slowly varying function of $p$ with 
$C \simeq 0.4$ for charm quarks at RHIC energies and $\sim$ 0.7 at LHC energies,
for $E \leq $ 5 -- 6 GeV.
This leads to a rather large drag of $\sim$ 1.6/fm at RHIC and $\sim$ 2.7/fm
at LHC on charm quarks at $\tau=\tau_i$, where $\tau _i=0.25$ fm/c is determined 
by the onset of the kinetic equilibrium~\cite{munshi2}.

This has a very important implication. Consider a charm
quark having an energy $E_i$ of the order of a few GeV at time $\tau_i$ 
in such an expanding and
chemically equilibrating plasma. Due to this large drag, the charm 
 quark produced initially will come to rest  very quickly and diffuse. 
We have verified this by performing numerical calculations of the final
momentum of charm quarks which propagate under such a drag. We find that,
irrespectively of the initial energy ($E_i \leq$ 5--6 GeV) the final energy of
the charm quark is about 1.5 -- 1.6 GeV, in the cases considered here.
Recall again that the charm quarks do not come to a stop
if only the collisional energy-loss is included~\cite{munshi}. 

Thus we conclude that the radiative energy-loss of heavy quarks produced
initially in relavisitic nuclear collisions plays a dominant role
in pulling them to a stop in the QGP both at RHIC and LHC energies.
Their final momentum distribution will then be determined by the temperature
at which the hadronization takes place. This could be the temperature
of the mixed phase, if such a phenomenon takes place.

We may add that Svetitsky and co-workers ~\cite{ben}
have actively investigated such a scenario. In their work the 
large drag coefficient arises due
to a large value of $\alpha_s \simeq 0.6$ and a fully equilibrated plasma,
even though only the collisional energy-loss is included.

Let us return to the discussion of the momentum distribution of charm quarks. 
(Similar considerations hold for bottom quarks.) It is expected that
the momentum distribution of charm quarks will be reflected in the
momentum distribution of charmed mesons, whose correlated decay will
provide a back-ground to dileptons from quark annihilation. We see
immediately that a look at the $p_{\bot}$ distribution of these leptons 
may help us to isolate the two contributions, as they should be
very different for the two sources. Shuryak~\cite{shur} and Lin
et al.~\cite{lin} have 
argued that the correlated charm decay back-ground for dileptons
may be suppressed if the energy-loss of charm quarks is taken as 1--2
GeV/fm. Our study lends a strong support to their conclusion
which were obtained by attributing an arbitrarily
assumed value for the energy-loss.

In conclusion, we have estimated the radiative energy-loss of heavy quarks
propagating in a quark gluon plasma. This, along with the (fairly small)
 collisional energy-loss acts as a strong drag force on heavy quarks,
 which pulls them 
to a stop even in a chemically equilibrating and cooling plasma.  This ensures
that the momenta of the resulting charm mesons will be determined
by the hadronization temperature. The correlated decay of such charm mesons
will then no longer pose a back-ground for dileptons having their 
origin in the quark-antiquark annihilation at least at large invariant mass. This 
separation could
even be made easier by measuring the $p_{\bot}$ distribtuion of the
lepton pairs. In a future publication we shall report the result of
the transverse hydrodynamic flow of the plasma on these conclusions
~\cite{munshi2}.

\bigskip
We gratefully acknowledge helpful discussions with Bikash Sinha.
One of us (D.K.S.) would like to thank for the hospitality of the University of Bielefeld,
where part of this work was done.  


\begin{thebibliography}{37}
\bibitem{correction} An estimate of radiative energy-loss of charm quarks 
was also given in Ref.~\cite{munshi}. The results given here were obtained after
more elaborate checks on the applicability of the approximation used by
Gunion and Bertsch~\cite{gunion} and putting $p'$ =3T while getting Eq.
(\ref{qtmax}). The results for radiative energy-loss
 plotted in figs. (2) of Ref.~\cite{munshi} unfortunately are  wrong
by a factor of $1/\hbar c \sim 5$ by which they should be multiplied. The 
results involving the collsional energy-loss given there are correct.
\end{thebibliography}
\end{document}